\documentclass[]{emulateapj}

\begin{document}

\newcommand{\Eiso}{E_{\rm iso}}
\newcommand{\Egamma}{E_{\gamma}}
\newcommand{\Egiso}{E_{\gamma,{\rm iso}}}
\newcommand{\ERiso}{E_{R,{\rm iso}}}
\newcommand{\FluR}{S_R}
\newcommand{\FluGamma}{S_\gamma}
\newcommand{\PLidx}{\alpha}
\newcommand{\PHidx}{\Gamma_{\rm ph}}
\newcommand{\NH}{N_H}
\newcommand{\cosmology}{$H_0=65$ km s$^{-1}$ Mpc$^{-1}$, $\Omega_\Lambda = 0.7$ and $\Omega_m = 0.3$}

\shorttitle{Explosive activity in GRB 060206}
\shortauthors{Wo\'zniak et al.}

\title{RAPTOR observations of delayed explosive activity in the high-redshift $\gamma$-ray burst GRB 060206}

\author{
P. R. Wo\'zniak,
W. T. Vestrand,
J. A. Wren,
R. R. White,
S. M. Evans,
and D. Casperson
}

\affil{Los Alamos National Laboratory, MS-D466, Los Alamos, NM 87545 \\
email: (wozniak, vestrand, jwren, rwhite, sevans, dcasperson)@lanl.gov}

\vspace{0.5cm}

\begin{abstract}
The RAPid Telescopes for Optical Response (RAPTOR) system at Los Alamos National Laboratory observed
GRB 060206 starting 48.1 minutes after $\gamma$-ray emission triggered the Burst Alert Telescope
(BAT) on-board the Swift satellite. The afterglow light curve measured by RAPTOR shows
a spectacular re-brightening by $\sim$1 mag about 1 h after the trigger and peaks at $R\sim16.4$ mag.
Shortly after the onset of the explosive re-brightening the OT doubled its flux on a time-scale
of about 4 minutes. The total $R$-band fluence received from GRB 060206 during this episode
is $2.3\times10^{-9}$ erg cm$^{-2}$. In the rest frame of the burst ($z$ = 4.045) this
yields an isotropic equivalent energy release of $\Eiso \sim 0.7\times10^{50}$ erg in just a narrow UV band
$\lambda \simeq 130\pm22$ nm. We discuss the implications of RAPTOR observations for
untriggered searches for fast optical transients and studies of GRB environments at high redshift.
\\
\end{abstract}


\section{Introduction}
\label{sec:intro}

Since the launch of the Swift satellite (\citealt{geh04}) in the fall of 2004, the number of Gamma-Ray Bursts (GRBs)
with known high redshifts
has been rapidly increasing. The sample of classical long-soft GRBs with $z>2.5$ includes at least 10 objects
(\citealt{jak06,jak05}). The current record holder, GRB 050904 at $z=6.295$ (\citealt{kaw05}), is located close
to the boundary of the reionized Universe (\citealt{wat06,tot05}), on par with the most distant galaxies and quasars
known today (e.g. \citealt{kod03,fan03}).

These developments are due to both Swift's sensitivity to high-$z$ events and high precision rapid localizations from
the BAT instrument (c.f. \citealt{ber05}). The immediate self-followup capability of Swift in X-ray and optical/UV bands
(\citealt{geh04}) combined with observations from fast-slewing robotic telescopes on the ground
(e.g. \citealt{ake03,blo04,boe04,cov04,gui05,per04,ves02}) greatly improves the probability of catching an optical
counterpart before it is too faint to grant a high resolution spectrum. An equally important outcome is well sampled,
multi-wavelength light curves that enable studies of broad-band spectral evolution reaching into the critical
first minutes of the explosion (e.g. \citealt{blu06,ves05,ves06,ryk05}).

Fast response technology played an important role in recent discoveries of the prompt optical/IR light in GRBs
(\citealt{ves05,ves06,bla05}), the X-ray flares (e.g. \citealt{obr06,bur05a,bur05b}), the rich structure
in temporal profiles of early afterglows with multiple breaks (e.g. \citealt{obr06,nou05}) and signatures of
the extended internal engine activity (e.g. \citealt{fal05,fal06}). These newly established features are providing
powerful clues about the physics of GRB explosions, their progenitors and surrounding environments. It is essential
to study those phenomena across a wide redshift range, as metallicity effects are likely to make distant GRBs
longer, more energetic, and easier to produce (\citealt{woo06a,woo06b}).

In this letter we present new evidence for late time explosive activity in high-redshift GRBs based on optical
observations of GRB 060206 ($z \simeq 4$) collected by the RAPTOR experiment.

\section{Observations}
\label{sec:data}

On 2006, February 6, 04:46:53 UT (trigger time; hereafter $t = 0$), the Burst Alert Telescope (BAT) instrument of the Swift satellite
(\citealt{geh04}) detected GRB 060206 (trigger number 180455). The temporal profile over the interval $t = [-1, 10]$ s
in the combined BAT energy range 15--350 keV is a single, bright gaussian-like peak with duration $T_{90} = 7\pm2$ s
(\citealt{mor06a,pal06}). The 15--350 keV fluence, the peak flux and the photon index of the time-averaged spectrum were
subsequently measured to be, respectively, $(8.4 \pm 0.4) \times10^{-7}$ erg cm$^{-2}$, 2.8 ph cm$^{-2}$ s$^{-1}$ and 1.06$\pm$0.34
(\citealt{pal06}). The on-board location (\citealt{mor06a}) was distributed in near-real time through the GRB Coordinates Network (GCN)
at 04:47:07.7 UT, $t \simeq 14.7$ s.

\begin{figure}
\vspace{4.5cm}
\includegraphics{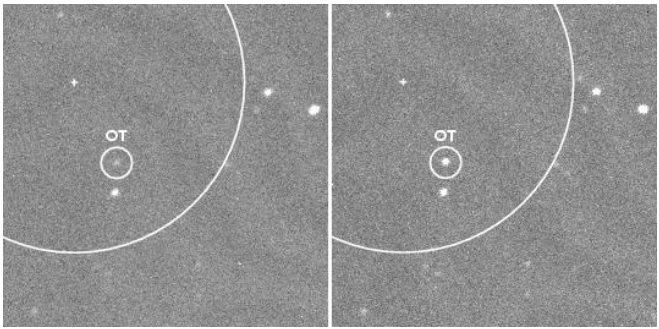}
\figcaption[]{\label{fig:frames} RAPTOR-S images of GRB 060206. The optical transient (OT) was a $R=17.3$ mag
object about 48.1 min after the trigger (left frame), and only 11 min later had increased its flux by almost 1 mag (right frame).
The initial BAT localization is marked with a cross. The error radius is 3$\arcmin$ (large circle).
}
\end{figure}

About $1$ s later the RAPTOR-S telescope received the BAT localization, which placed the object below the altitude limit
of the instrument ($\sim$19$\arcdeg$). The RAPTOR scheduling software restarted the alert response sequence at 05:34:58 UT,
$t = 48.1$ min, i.e. immediately after the BAT position became accessible. The system collected a series of 100 30-second
images covering the next hour. Despite high airmass and windy conditions, the final quality of our unfiltered images
is very good (Fig.~\ref{fig:frames}). RAPTOR-S is an 0.4-m, fully autonomous robotic telescope, typically operated at focal
ratio f/5. It is equipped with a 1k $\times$ 1k pixel CCD camera employing a back-illuminated Marconi CCD47-10 chip
with 13 $\mu$ pixels. The telescope is owned by Los Alamos National Laboratory and located at the Fenton Hill Observatory
(106.67$\arcdeg$ W, 35.88$\arcdeg$ N) at an altitude of $\sim$2500 m in the Jemez Mountains of New Mexico.

The RAPTOR data processing system has capability to automatically locate optical counterparts to GRBs in real time
(\citealt{woz06a}). About 5 min into the response sequence for GRB 060206, i.e. after nine images,
the system identified an uncatalogued 17-th magnitude variable object at RA = 13:31:43.44, DEC = 35:03:03.2 (J2000), a location
$\sim$1.6$\arcmin$ from the on-board BAT position and well inside the 3$\arcmin$ error radius. This machine-generated identification
of a possible optical transient (OT) associated with GRB 060206 was forwarded immediately (at 05:39:50 UT) to the RAPTOR rapid
response team using a wireless network. A quick examination of the unfolding light curve indicated that 50 min after the trigger
the candidate OT was undergoing a dramatic re-brightening. The initial report of a non-detection in a 72-second $V$-band frame taken
by UVOT at $t \simeq 58$ s (\citealt{mor06a}) seemed to contradict the case for a long lasting optical event.
This unusual behavior of the OT prompted our further investigation of its association with GRB 060206 before announcing
the optical counterpart.

In the mean time, Fynbo et al. (\citealt{fyn06a}) reported a candidate afterglow found using ALFOSC instrument
of the 2.5-m Nordic Optical Telescope (NOT) on La Palma. The position of the proposed afterglow was consistent with that
of the variable source found by RAPTOR. The afterglow hypothesis was quickly confirmed by re-examination of the UVOT data
(\citealt{boy06}) and determination of redshift $z = 4.045$ for the candidate from NOT spectra (\citealt{fyn06b,fyn06c}).
At this time, the RAPTOR light curve clearly showed that GRB 060206 displayed a spectacular re-brightening to $\sim$16.3 mag
around $t=1$ h and had resumed its fading behavior (\citealt{woz06b}). The re-brightening was promptly confirmed by observations
with the 2-m Liverpool telescope (\citealt{gui06}).

\begin{figure}
\vspace{6.65cm}
\includegraphics{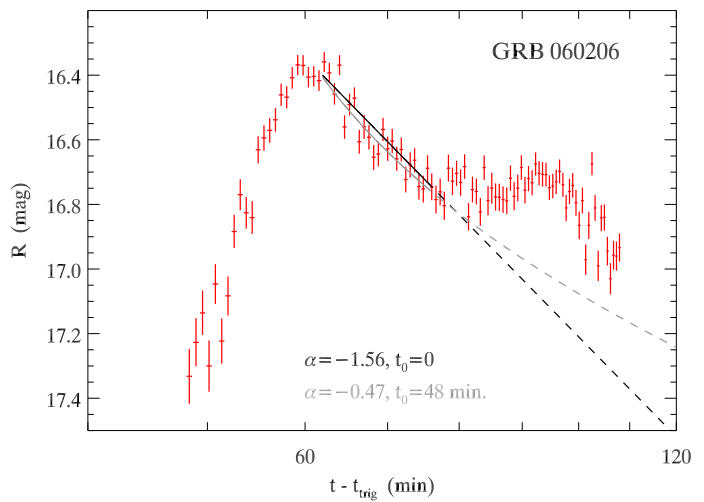}
\figcaption[]{\label{fig:lc} RAPTOR-S optical light curve of GRB 060206. The two power-law models shown have about the same
goodness of fit in the time interval 62--76 min (\S\,\ref{sec:results}). At the onset of the explosive
re-brightening the OT flux doubled on a time-scale of 4 minutes. The fading behavior of the OT
is consistent with the power-law flux decay of index $\alpha=-1$.
}
\end{figure}

\section{Photometry}
\label{sec:redux}

After standard corrections for bias, dark current and flat field responses, the RAPTOR real time photometric
pipeline performs source detection and centroiding using custom routines based on algorithms similar to those
used in the SExtractor package (\citealt{ber96}). Images are smoothed with a Gaussian kernel with FWHM = 4.0 pixels
for the purpose of detecting flux peaks. This is appropriate for a typical (largely instrumental) seeing of $\sim$5\arcsec
(FWHM) and 1.2\arcsec\,pixel$^{-1}$ plate scale of RAPTOR-S. Object magnitudes are measured using simple
aperture photometry within a 5-pixel radius of the centroid. Errors are estimated from propagation of the photon noise.

The flux scale with about 3\% internal consistency was established using $\sim$30 high S/N objects covering entire
$24\arcmin\times24\arcmin$ field of view. Our unfiltered optical band has the effective wavelength close to that
of the standard $R$ band, but it has a larger width. For lack of the instrumental color information, one is forced
to assume that all objects have the color of a mean star in the field. All measurements were calibrated
to standard $R$-band magnitudes using 26 sources from the USNO-B1.0 catalog. Comparison point sources were selected to have
galaxy-star separation index GS $\geq$ 5 and $R2$ magnitudes brighter than 18.0 in the catalog (\citealt{mon03}).

\section{Results}
\label{sec:results}

The final RAPTOR photometry of the optical afterglow of GRB 060206 expressed on the $R$-band scale is given
in Table~\ref{tab:data} and plotted in Fig.~\ref{fig:lc}. During the first $\sim$700 s after the burst position
became visible ($t = 48$--60 min) the OT flux is sharply rising from $R\sim17.3$ to a peak value $\sim$16.4 mag.
The peak flux was roughly maintained for the next $\sim$4 minutes. The subsequent decline to $\sim$16.75 mag at $t=80$ min
was followed by a secondary $\sim$0.1 mag brightening around $t=95$ min. For the remaining 15 min of the response
sequence the OT was somewhat erratically fading to $\sim$17.0 mag. The majority of small scale variations in the light
curve are simply statistical fluctuations due to measurement uncertainties. However, a visual inspection of the
RAPTOR images indicates that some of the larger changes on time-scales of a few minutes are intrinsic to the OT.
We estimated the rate of flux decay after the first brightening episode (time interval 62--76 min) by fitting a power-law model
$f(t) \propto f(t_d)[(t-t_0)/(t_d-t_0)]^\alpha$, where $t_d$ is the beginning of the decline and $t_0$ is typically associated
with the explosive energy injection. The best fit model with $t_0 = 0$ has $\alpha=-1.56\pm0.04$, and assuming $t_0=48$ min
we find $\alpha=-0.47\pm0.03$ (Fig.~\ref{fig:lc}). While the latter model is a better fit ($\chi^2/d.o.f.=34.6/21$)
compared to $\chi^2/d.o.f.=37.7/21$), the difference is hardly significant and both models are formally unacceptable.
The fading behavior of the OT is roughly consistent with the power-law flux decay of index $\alpha=-1$.

The remarkable late time activity present in GRB 060206 has never been observed before in the optical energy range.
About 1 h into the event we measured the peak rate of flux increase 0.2 mag min$^{-1}$ (flux doubling time-scale of $\sim$4 min)
and a lower limit of $\sim$1 mag for the total flux increase in 12 minutes.
The estimated total fluence in $R$-band
photons received by RAPTOR-S during its $\sim$1 h response is $2.3\times10^{-9}$ erg cm$^{-2}$. Note, however, that
for a $z=4.045$ object the rest frame energies of optical photons detected by RAPTOR correspond to the far UV regime,
$\lambda \sim 130$ nm. The isotropic equivalent energy emitted by GRB 060206 around $\lambda \simeq 130\pm22$ nm
is $\Eiso \sim 0.7\times10^{50}$ erg (assuming flat cosmology with \cosmology). In \S\,\ref{sec:discussion}
we discuss broader ramifications of these observations.

\begin{figure}
\vspace{6.65cm}
\includegraphics{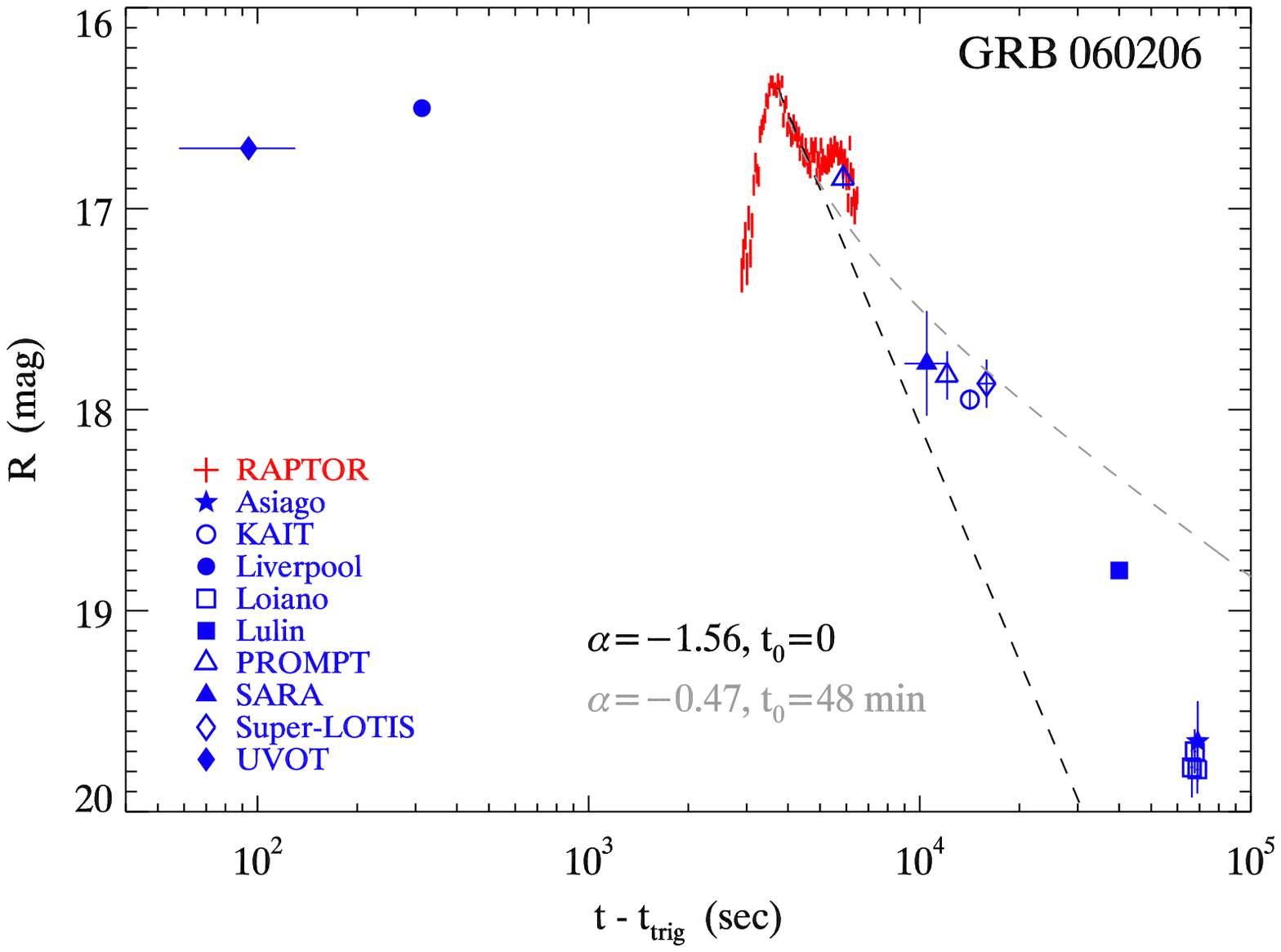}
\figcaption[]{\label{fig:comb_lc} RAPTOR optical light curve of GRB 060206 compared
with measurements from other instruments:
Asiago (\citealt{mal06}), KAIT (\citealt{li06}), Liverpool (\citealt{gui06}), Loiano (\citealt{gre06}), Lulin (\citealt{lin06}),
PROMPT (\citealt{hai06}), SARA (\citealt{hom06}), Super-LOTIS (\citealt{mil06}).
All measurements were reported as $R$ magnitudes except for PROMPT ($r'$) and UVOT ($V$).
The lines are model fits described in \S\,\ref{sec:results}.
}
\end{figure}

\section{Discussion}
\label{sec:discussion}

In Fig.~\ref{fig:comb_lc} we compare the RAPTOR light curve with OT measurements from several other instruments.
All measurements were taken at face value, i.e. without any corrections for slight differences due to photometric
filters and comparison catalogs. From the UVOT $V$-band measurement at $t=58$ s (\citealt{boy06}) and the report
of fading behavior after $t=5.23$ min (\citealt{gui06}) we can infer that the OT peaked at least once
in the time interval preceding the RAPTOR observations. However, Fig.~\ref{fig:comb_lc} suggests that
the energy radiated by the OT during the re-brightening around $t=1$ h is a substantial (if not dominant)
part of the energy budget in the observed $R$ band.

Pronounced re-brightenings in GRB afterglows may arise
due to strongly non-uniform density profile of the material in the path of the blast wave (e.g. \citealt{tam05}).
Another possibility is that the complexity of the pulse train is due to the internal engine itself and the energy
is injected at a variable rate over time interval comparable to the time-scale on which the shocked medium
responds (see \cite{pro06} for a recent discussion).
For the best observed GRB afterglows we are at the verge of performing a proper reverberation
mapping (\citealt{ves06,ves05}) using the prompt emission as a tracer of the input energy and shock development
models as a fiducial response function (e.g. \citealt{mes97,sar99}). Panchromatic light curves can then constrain
the models because the derived density profile should be independent of the observed energy range.

\subsection{Comparison to GRB 050904}
\label{sec:comp}

The most distant $\gamma$-ray burst know today, GRB 050904 ($z=6.295$), was well observed.
Therefore, we can make a number of interesting comparisons between the multi-wavelength
properties of GRB 060206 and the current record holder for the distance. The duration
of GRB 050904 in $\gamma$-rays was $T_{90} \simeq 225$ s, compared to $T_{90} \simeq 7$ s
for GRB 060206\footnotemark. The difference is large even when transformed to the rest frame of the burst.
GRB 050904 was an average size event with $\gamma$-ray fluence
$\FluGamma = (5.1\pm0.2)\times10^{-6}$ erg cm$^{-2}$ and $\Egiso = 3.8\times10^{53}$ erg (\citealt{sak05}).
In constrast GRB 060206 produced roughly an order of magnitude less energy: $\FluGamma = (8.4\pm0.4)\times10^{-7}$
erg cm$^{-2}$ and $\Egiso = 3.1\times10^{52}$ erg (\citealt{pal06}).

\footnotetext{At redshift $z=1$ GRB 060206 would qualify as a short burst
($T_{90} < 2$ s), although this could be an artifact of low S/N ratio.}

The XRT spectrum of the early X-ray afterglow in GRB 050904 is harder than that of GRB 060206
with $\PHidx \sim 1.2$ (\citealt{wat06}) and $\PHidx \sim 2.0$ (\citealt{mor06b}) respectively.
GRB 050904 displayed high amplitude flaring activity lasting several hours in the X-ray band (e.g. \citealt{wat06})
with features not unlike the optical/UV flare in GRB 060206 observed by RAPTOR. Although the burst frame peak optical/UV flux
of GRB 060206 measured by RAPTOR (c.f. \citealt{gui06}) appears fainter by $\sim$2 mag when compared to GRB 050904
(\citealt{boe05}), the latter event emitted less energy in optical/UV photons during the afterglow phase relative
to the $\gamma$-ray output.

\subsection{IGM and circumburst medium at $z=4$}
\label{sec:absorption}

GRBs have X-ray and optical spectra with featureless continua that are easy to model. 
It turns out that they are also sufficiently long-lived and bright to be used in studies of
the inter-galactic medium and the immediate GRB neighborhood at high redshifts. \cite{wat06} conclude that
in the X-ray domain GRB afterglows are more promising beacons than AGNs. The number of bright optical
flashes from high-$z$ GRBs detected so far suggests that this is also true for high-resolution optical spectroscopy.

From spectral model fits in the 0.5--20 keV energy range of the XRT detector \cite{wat06} found evidence
for low energy absorption toward GRB 050904 with column densities $N_H \sim 8\times10^{20}$ cm$^{-2}$ in excess
of the Galactic value for this line of sight. Excess absorption of the order $3\times10^{20}$ cm$^{-2}$
(although with a large uncertainty) was reported for GRB 060206 based on similar spectral models (\citealt{mor06b}).

The column density of neutral hydrogen inferred from a particular level of attenuation in soft X-rays
strongly depends on the assumed redshift for the absorber. A modest difference at $z=0$ translates
into large column densities for any intrinsic absorption in high-redshift objects. In GRB 050904
\cite{wat06} found $N_H \sim 3\times10^{22}$ cm$^{-2}$, and in GRB 060206 $N_H\sim10^{22}$ cm$^{-2}$
is implied, assuming a high-$z$ absorber with a covering factor of 1. For a typical dust-to-gas ratio in the Galaxy there is
about 5 mag of visual extinction $A_V$ for every $10^{22}$ HI atoms cm$^{-2}$. Therefore, a naive estimate
of $A_V$ due to material around GRBs 050904 and 060206 is 5--15 mag. UV extinction is several times larger,
even in low metallicity environments like SMC. This amount of obscuration would hide any realistic
level of optical/UV emission in high-$z$ events. Lowering the metal abundances assumed
by spectral fits would only increase the amount of neutral hydrogen implied by the best fit model.
Unless the inferred HI columns are an artifact of the uncertain low energy XRT calibrations, one of the
above assumptions is invalid. Fine scale structure of the absorber in our Galaxy may explain
the apparent excess absorption. In the future, a careful analysis
of the X-ray/UV/optical absorption in high-redshift GRBs should provide sensitive diagnostics of the
element composition, ionization state and clumpiness of the medium surrounding the GRB progenitor.

\subsection{Implications for orphan GRB searches}
\label{sec:orphans}

GRB 060206 was an ``optically rich'' burst.
An interesting possibility is that the optical/UV radiation emitted during the re-brightening
of the afterglow originates in the jet material traveling at a relatively low Lorentz factor $10 < \Gamma <100$,
and producing relatively little $\gamma$-ray emission. This situation could arise in any of the scenarios involving
a highly structured or patchy jet viewed slightly off-axis (\citealt{nak03}), a ``dirty fireball'' with strong baryon loading
(e.g. \citealt{rho03,hua02}), or variations in the Lorentz factor of the ejecta propelled by the central engine
(e.g. \citealt{ree98,pan05}).
Therefore, observations of GRB 060206 support the case for existence of events in which the optical/UV emission
would completely dominate or perhaps precede the actual $\gamma$-ray burst. In any case, it is a fact that optical emission
lasting tens of minutes can be detected with a modest wide-field telescope in GRBs occurring over a large volume of the Universe.
A well designed untriggered search using small optical telescopes is likely to be successful and could significantly
expand the range of GRB parameters covered by observations.
\\

\acknowledgements

This research was performed as part of the Thinking Telescopes and RAPTOR projects supported by
the Laboratory Directed Research and Development (LDRD) program at LANL.

\begin{deluxetable}{rcccrcccrcccrcc}
\tablewidth{16cm}
\tablecaption{\label{tab:data}{RAPTOR photometry of GRB 060206.\tablenotemark{a}}}
\tablehead{
\colhead{$t_{\rm start}$} &
\colhead{$R$} &
\colhead{$\sigma$} &
\colhead{~~} &
\colhead{$t_{\rm start}$} &
\colhead{$R$} &
\colhead{$\sigma$} &
\colhead{~~} &
\colhead{$t_{\rm start}$} &
\colhead{$R$} &
\colhead{$\sigma$} &
\colhead{~~} &
\colhead{$t_{\rm start}$} &
\colhead{$R$} &
\colhead{$\sigma$} \\
\colhead{\makebox[1.0cm]{(min)}} &
\colhead{\makebox[1.0cm]{(mag)}} &
\colhead{\makebox[1.0cm]{(mag)}} &
\colhead{} &
\colhead{\makebox[1.0cm]{(min)}} &
\colhead{\makebox[1.0cm]{(mag)}} &
\colhead{\makebox[1.0cm]{(mag)}} &
\colhead{} &
\colhead{\makebox[1.0cm]{(min)}} &
\colhead{\makebox[1.0cm]{(mag)}} &
\colhead{\makebox[1.0cm]{(mag)}} &
\colhead{} &
\colhead{\makebox[1.0cm]{(min)}} &
\colhead{\makebox[1.0cm]{(mag)}} &
\colhead{\makebox[1.0cm]{(mag)}}
}
\startdata
   48.1 &   17.33 &   0.09 & &     63.1 &   16.46 &   0.03 & &     78.2 &   16.69 &   0.04 & &     93.2 &   16.71 &   0.04 \\
   48.7 &   17.23 &   0.07 & &     63.7 &   16.37 &   0.03 & &     78.8 &   16.73 &   0.04 & &     93.8 &   16.71 &   0.04 \\
   49.3 &   17.14 &   0.07 & &     64.3 &   16.56 &   0.04 & &     79.4 &   16.70 &   0.04 & &     94.4 &   16.75 &   0.04 \\
   49.9 &   17.30 &   0.08 & &     64.9 &   16.49 &   0.03 & &     80.0 &   16.73 &   0.04 & &     95.0 &   16.74 &   0.04 \\
   50.5 &   17.05 &   0.06 & &     65.5 &   16.47 &   0.03 & &     80.6 &   16.68 &   0.04 & &     95.6 &   16.73 &   0.04 \\
   51.1 &   17.22 &   0.07 & &     66.1 &   16.61 &   0.04 & &     81.2 &   16.84 &   0.04 & &     96.2 &   16.70 &   0.04 \\
   51.7 &   17.08 &   0.06 & &     66.7 &   16.56 &   0.04 & &     81.8 &   16.75 &   0.04 & &     96.8 &   16.74 &   0.04 \\
   52.3 &   16.88 &   0.05 & &     67.3 &   16.59 &   0.04 & &     82.4 &   16.76 &   0.04 & &     97.4 &   16.81 &   0.04 \\
   52.9 &   16.77 &   0.05 & &     67.9 &   16.65 &   0.04 & &     83.0 &   16.82 &   0.04 & &     98.0 &   16.76 &   0.04 \\
   53.5 &   16.83 &   0.05 & &     68.5 &   16.64 &   0.04 & &     83.6 &   16.69 &   0.04 & &     98.6 &   16.74 &   0.04 \\
   54.1 &   16.84 &   0.05 & &     69.2 &   16.57 &   0.04 & &     84.2 &   16.79 &   0.05 & &     99.2 &   16.80 &   0.04 \\
   54.7 &   16.63 &   0.04 & &     69.8 &   16.63 &   0.04 & &     84.8 &   16.75 &   0.05 & &     99.9 &   16.87 &   0.04 \\
   55.3 &   16.59 &   0.04 & &     70.4 &   16.60 &   0.04 & &     85.4 &   16.78 &   0.04 & &    100.5 &   16.79 &   0.04 \\
   55.9 &   16.57 &   0.04 & &     71.0 &   16.66 &   0.04 & &     86.0 &   16.78 &   0.04 & &    101.1 &   16.97 &   0.05 \\
   56.5 &   16.54 &   0.04 & &     71.6 &   16.63 &   0.04 & &     86.6 &   16.79 &   0.04 & &    101.7 &   16.87 &   0.04 \\
   57.1 &   16.46 &   0.03 & &     72.2 &   16.72 &   0.04 & &     87.2 &   16.79 &   0.04 & &    102.3 &   16.67 &   0.04 \\
   57.7 &   16.47 &   0.03 & &     72.8 &   16.68 &   0.04 & &     87.8 &   16.72 &   0.04 & &    102.9 &   16.81 &   0.04 \\
   58.3 &   16.41 &   0.03 & &     73.4 &   16.66 &   0.04 & &     88.4 &   16.77 &   0.04 & &    103.5 &   16.99 &   0.05 \\
   58.9 &   16.37 &   0.03 & &     74.0 &   16.75 &   0.04 & &     89.0 &   16.75 &   0.04 & &    104.1 &   16.84 &   0.04 \\
   59.5 &   16.37 &   0.03 & &     74.6 &   16.75 &   0.04 & &     89.6 &   16.69 &   0.04 & &    104.7 &   16.84 &   0.04 \\
   60.1 &   16.41 &   0.03 & &     75.2 &   16.69 &   0.04 & &     90.2 &   16.76 &   0.04 & &    105.3 &   16.94 &   0.04 \\
   60.7 &   16.40 &   0.03 & &     75.8 &   16.75 &   0.04 & &     90.8 &   16.72 &   0.04 & &    105.9 &   17.03 &   0.05 \\
   61.3 &   16.42 &   0.03 & &     76.4 &   16.78 &   0.04 & &     91.4 &   16.73 &   0.04 & &    106.5 &   16.96 &   0.04 \\
   61.9 &   16.36 &   0.03 & &     77.0 &   16.76 &   0.04 & &     92.0 &   16.67 &   0.04 & &    107.1 &   16.96 &   0.05 \\
   62.5 &   16.39 &   0.03 & &     77.6 &   16.80 &   0.04 & &     92.6 &   16.70 &   0.04 & &    107.7 &   16.93 &   0.04 \\
\enddata

\tablenotetext{a}{All measurements were obtained with the RAPTOR-S instrument. Our unfiltered magnitudes were transformed
to $R$-band scale using field stars from USNO-B1.0 catalog, and were not corrected for extinction
(Galactic $E(B-V)$ reddening is only 0.018 mag; \citealt{sch98}).}

\end{deluxetable}

\end{document}